\documentstyle[12pt]{article}
\begin{document}
\baselineskip=20pt

\title{\bf Nuclear Effects on Bremsstrahlung Neutrino Rates of Astrophysical Interest}

\author{Sabin Stoica\\
\it Horia Hulubei National Institute of Physics and
Nuclear\\
Engineering, P.O. Box MG-6, 76900 Bucharest-Magurele, Romania\\
J.E. Horvath\\
\it IAG/USP\\
Av. M. St\'efano 4200, Agua Funda, 04301-904 S\~ao Paulo SP, Brazil\\}
\maketitle
\pagestyle{empty}
\vskip5mm

\noindent
{\bf Abstract} 

We calculate in this work the rates for the neutrino 
pair production by nucleon-nucleon 
bremsstrahlung taking into account the full 
contribution from  a nuclear one-pion-exchange potential. 
It is shown that if the temperatures are low enough ($T \leq 20 MeV$), 
the integration over the nuclear part can be done for the
general case, ranging from the completely degenerate (D) to the  
non-degenerate (ND)
regime. We find that the inclusion of the full nuclear  
contribution enhances the neutrino pair production by 
$nn$ and $pp$ bremsstrahlung by a factor of about two 
in both the D and ND limits when compared with previous calculations.
This result may be relevant for the 
physical conditions of interest in the semitransparent regions near the 
neutrinosphere in type II supernovae, cooling of neutron stars 
and other astrophysical situations.  

\section{Introduction}

Neutrino production processes in type II supernovae (SNII) and 
proto-neutron stars has 
received recently a considerable boost because of their importance in 
understanding of a variety of interesting phenomena like the very explosion 
mechanism, aspects related to the SNII nucleosynthesis, the interpretation of the
neutrino signal from SN, the thermalization and cooling mechanisms, among others.

Several physical processes producing neutrinos have been investigated. The 
most important ones are electron scattering, electron-positron annihilation, 
pion-nucleon and NN  scattering, NN  bremsstrahlung, plasmon- and
photo-neutrino production. Their relative importance for different stages of the early 
proto-neutron star evolution (including the explosion itself) 
is a consequence of quite different functional dependences on $\rho, T$. An analogous 
statement can be carried over to the great deal of work done on the cooling of 
mature neutron stars \cite{[TS00]}.

One of the important questions in supernova physics 
is the actual post-bounce evolution driven by neutrino 
emission, thought to be crucial for the long-term delayed shock revival driven by 
neutrinos (the Wilson mechanism). Neutrino production 
and thermalization were, for a long time, 
assumed to be dominated by pair annihilation and electron scattering respectively; 
at least for the $\nu_\mu$ and $\nu_\tau$ species (given that 
charged current reactions are certainly the main source for $\nu_e$'s).
Recenty, however, increasingly 
accurate calculations revealed that the energy transfer rate in 
neutrino ($\nu_{\mu}$) - nucleon scattering may be up to one order of magnitude 
larger than the previous estimations, making this process competitive with 
the $\nu_{\mu}$-electron scattering as an equilibration mechanism \cite{[BS98]},
\cite{[HR98]}, \cite{[SIG97]}. Also, similar calculations for the total neutrino
emissivity from NN bremsstrahlung indicate that this process may 
compete with $e^+ e^-$ annihilation \cite{[HR98]}, \cite{[BUR20]},
\cite{[FSB75]} as a source for the $\nu_\mu$ and $\nu_\tau$ neutrinos. 
Bremsstrahlung rates have also been revisited for those non-degenerate to 
semidegenerate conditions, including the potentially important $np$ channel.
As a general feature of nucleon scattering, NN bremsstrahlung and URCA processes, 
it has proved quite difficult to treat the strong NN interaction matrix elements 
which are responsible for the $\nu$ production together with the weak interactions.

In the early calculations
this interaction has been treated either by computing the overlap integrals
associated with the initial and final nucleon wave functions \cite{[BW65]}
or through the use of a Fermi liquid parametrization \cite{[FSB75]}. Later on, 
a NN potential based on the one-pion-exchange (OPE) approximation has been
deduced by several authors \cite{[FM79]}, \cite{[HR98]}, \cite{[HR95]},
\cite{[RSS96]}, \cite{[IWA84]}, \cite{[BT88]}. However, in all these papers the 
full momentum transfer dependence of the 
OPE potential has been approximated (being, for instance
replaced by the Fermi momentum) or even the whole square of the matrix element 
taken as a constant  \cite{[BT88]}, \cite{[TBH20]}. 
However, for an accurate computation of these
processes (and thus a realistic computation of the post-bounce evolution of 
type II supernovae) an accurate treatment of the nuclear potential is certainly required. 

We present in this work a method for treating explicitly the momentum dependence
of the OPE potential involved in the calculation of the neutrino pair
production from NN bremsstrahlung. Actually, the method is rather general and 
could be also used for the treatment of other processes involving NN potential. 
We show that in particular physical conditions, characterized by
temperatures $T  \, <  m_{\pi}^{2}/m \sim \,  20 MeV$, the integral over the nuclear matrix 
element collapses, in 
a good approximation, to an integral which is independent of angles. The
remaining part of the ME contribution, which depends only on the nucleon and
neutrino energies, can be easily integrated numerically. Some results of this evaluation 
for non-degenerate and degenerate nucleons are presented.

\section{Calculations}
 
The total volumetric emissivity for the neutrino pair production by nucleon 
bremsstrahlung is given by

$$ Q_{\nu} = \frac{2\pi}{\hbar} {\int \left[\Pi_1^4 \frac{d^3{\bf p}_i} 
{(2 \pi)^3}\right] \frac{ d^3{\bf q}_{\nu}}{(2 \pi)^3 2 \omega_1}\frac{d^3
{\bf q}_{\bar{\nu}}}{(2\pi)^3 2\omega_2}\omega \left({\it s} \Sigma {\vert
M \vert}^2 \right) (2\pi)^3 \delta^4(P){\it F(f)}} \eqno(1)$$
\noindent
where 

$${\it F(f)} = {\it f_1 f_2 (1-f_3) (1-f_4)}
\eqno(2)$$
\noindent
is the product of  Fermi functions for the initial (1,2) and final (3,4) nucleons 
$ {\it f_i} = \left(\exp^{\frac{E_i - \mu_i}{T}} + 1 \right)^{-1}$.
In eq.(1) ${\bf p}_i,~ i=1,4$  and ${\bf q}_{\nu}, {\bf q}_{\bar {\nu}}$ are the 
nucleon and neutrino momenta, respectively; 
$\omega=\omega_1 + \omega_2$ are the neutrino
energies ; {\it s}  is a symmetry factor taking into account the symmetry
of identical particles ({\it s} = 1/4 for $nn$ and $pp$, and {\it s}= 1 for the $np$ 
channel) and $E_i$ and $\mu_i$ are the energies and chemical potentials of the nucleons.  

In the non-relativistic limit $E_i \sim m + \frac{{\bf p}_i^2}{2m}$ and one
defines the chemical potential $\hat{\mu}=\mu - m$. 
With the introduction the non-dimensional quantities \cite{[BT88]}
$ y=\frac{\hat{\mu}}{T}\hskip1cm ; u_i = \frac{{\bf p}_i^2}{2mT}$
we write ${\it f_i} = \left(\exp^{u_i - y_i} + 1\right)^{-1}$.
The degenerate (D) limit is achieved for $y >> 1$, while in the 
non-degenerate (ND) limit $y << 1$.

In the OPE approximation the spin-summed NN
matrix element (ME) for $nn$ and $pp$ interactions has been derived by several 
authors (see for instance \cite{[HR98]}) and reads

$$ {\it s}~ \Sigma_{spins}{\vert M_{nn}\vert^2} = \hskip10cm $$
$$ = \frac{1}{4} \frac{ 64 G^2
g_A^2}{\omega^2}\left(\frac{f}{m_{\pi}}\right)^4 \left[
 \left(\frac{{\bf k}^2}{{\bf k}^2+m_{\pi}^2}\right)^2 + \left(\frac{{\bf l}^2}
{{\bf l}^2+m_{\pi}^2}\right)^2 + \frac{{\bf k}^2 {\bf l}^2}{({\bf k}^2 + 
m_{\pi}^2)({\bf l}^2 + m_{\pi}^2)}\right]\times$$ 
$$\times (\omega_1 \omega_2 -{\bf q}_{\nu}{\bf k}{\bf q}_{\bar{\nu}}{\bf k}) \equiv  
A \frac{\omega_1 \omega_2}{\omega^2} {\it M_{nn}} \eqno(3) $$
\noindent 

where $g_A = 1.26, f \simeq 1, A= 64 G^2 g_A^2
\left(\frac{f}{m_{\pi}}\right)^4$; ${\bf k} = {\bf p}_1 - {\bf p}_3 $ 
and ${\bf l} = {\bf p}_1 - {\bf p}_4 $ are the nucleon direct and exchange 
transfer momenta, respectively. In the non-relativistic approximation 
one can also neglect the neutrino momenta in eq.(3), since it is always 
small compared to 
the nucleon ones. For the sake of completeness we quote the corresponding 
ME for the $np$ interaction \cite{[HR98]}

$$ {\it s}~ \Sigma_{spins}{ \vert M_{nn} \vert ^2} = \hskip10cm$$
$$=\frac{1}{4} \frac{ 64 G^2 g_A^2}{\omega^2}
\left(\frac{f}{m_{\pi}}\right)^4 \left[
 \left(\frac{{\bf k}^2}{{\bf k}^2+m_{\pi}^2}\right)^2 + 2 \left(
\frac{{\bf l}^2}{{\bf l}^2 + m_{\pi}^2}\right)^2 + 
2 \frac{{\bf k}^2 {\b l}^2}{({\bf k}^2+
m_{\pi}^2)({\bf l}^2 + m_{\pi}^2)}\right]
(\omega_1 \omega_2 -{\bf q}_{\nu}{\bf k q}_{\bar{\nu}}
{\bf k}) \times$$
$$ \equiv A \frac{\omega_1 \omega_2}{\omega^2} \left[ 2{\it M_{nn}} - 
\left(\frac{{\bf k}^2}{{\bf k}^2+m_{\pi}^2}\right)^2\right]  
\eqno(4) $$

It is convenient to 
perform the integrals in the center-of-mass
coordinate system and define the new variables introduced by 
Brinkmann and Turner \cite{[BT88]}

$$ {\bf p}_+ = \frac{{\bf p}_1 + {\bf p}_2}{2};~~{\bf p}_- = \frac{{\bf p}_1 -
{\bf p}_2}{2};~~{\bf p}_{3c} = {\bf p}_3 + {\bf p}_+;~~{\bf p}_{4c} = 
{\bf p}_4 + {\bf p}_+ \eqno(5) $$ 

$$ cos{\gamma_1} = \frac{{\bf p}_+ {\bf p}_-}{\vert {\bf p}_+\vert \vert 
{\bf p}_-};~~cos{\gamma_c} = \frac{{\bf p}_+ {\bf p}_{3c}}{\vert {\bf p}_+
\vert \vert {\bf p}_{3c}};~~
cos{\gamma} = \frac{{\bf p}_- {\bf p}_{3c}}{\vert {\bf p}_-\vert \vert {\bf
p}_{3c}\vert} \eqno(6) $$

From the definition of the $u$ variables above, 
one can easily deduce the following relations 

$$ u_{1,2} = u_+ + u_- \pm 2(u_+ u_-)^{1/2}~~ ;~~ u_{1,2} = u_+ + u_{3c}
 \pm 2(u_+u_{3c})^{1/2} \eqno(7) $$ 

We now address the $nn$ and $pp$ expresion of the OPE potential and calculate
the corresponding emissivity. After some  
algebra one can express the  matrix element eq.(3) in terms of  the scalar 
combinations $a_s = {\bf k}^2 + {\bf l}^2$ and $a_p = {\bf k}^2  {\bf l}^2$ as 

$$ {\it M_{nn}} =  \left(3 - m_{\pi}^2 \cdot \frac{a_s(3a_p+2m_{\pi}^2a_s) +
m_{\pi}^2(7a_p+6m_{\pi}^2a_s)+3m_{\pi}^6}{a_p(a_p+2m_{\pi}^2a_s)+m_{\pi}^4(a_s^2+2a_p)+
2m_{\pi}^6a_s+m_{\pi}^8}\right) \equiv \,  \left(3 - M^{corr}_{nn}\right) \eqno(8) $$

\noindent 
Thus, the contribution of the nuclear ME can be splitted into a 
(already calculated) constant term and a correction to be evaluated.
It is worth mentioning that the main part of the nuclear potential contribution
(the constant term) is just 3, i.e. the limit to which the expression eq.(3) 
converges when the pion mass is neglected compared to the nucleon momentum transfer. 
The $a_p$ and $a_s$ terms have the explicit expressions 
$a_s = 2d(x +1/2 z) ~~~;~~~ a_p = d^2[x (x+z)sin^2{\gamma}+1/4z^2]$, 
with $d=2mT$, $x = 2u_{3c}$ and $z = \frac{\omega}{T}$. Again, after some 
algebra the correction part is cast in the form

$$ M^{corr}_{nn} = m_{\pi}^2 \times \frac{A_1 - A_2(c_g + s_g cos {\phi})^2}
{B_1 - B_2(c_g + s_g cos {\phi})^2 + B_3(c_g + s_g cos {\phi})^4} \eqno(9)$$
 
where $ c_g = cos{\gamma_1}\cdot cos{\gamma_c}~;~ s_g = sin{\gamma_1}\cdot
sin{\gamma_c}$ and $\phi$ is the difference between 
the azithmutal angles corresponding to $\gamma_1$ and $\gamma_c$ 
(in spherical coordinates). 

The coefficients $A_i$ and $B_j$ are polynomials of degree eight in the product
$(m_{\pi}^2mT)$,  depending  only on variables $x$ and $z$ but 
otherwise independent of the angles. 
$B_3$ is also binomial having a coefficient proportional to $(2mT)^4$, while 
$B_1$ and $B_2$ contain terms proportional to
$m_{\pi}^4$, $m_{\pi}^6$ and $m_{\pi}^8$. Therefore, whenever 
$mT \, < \, m_{\pi}^2$ (or equivalent, $T  \, < \, 20~ MeV$), $B_3$ 
can be neglected because its smallness compared to the other terms. 
This approximation can also be checked
numerically. The results of integration over $M_{nn}^{corr}$ with and without
the term $B_3$ differ each other by a very small amount, (less than 2\%) and 
greatly simplifies the calculation . The remaining integral over $\phi$ 
can be performed analytically yielding

$$ I_{\phi} = 2\pi \frac{A_2}{B_2} = m_{\pi}^2 \frac{3d(2x+z) + 
7m_{\pi}^2}
{2(d^2(x+z) + 4m^2T^2z^2 + d m_{\pi}^2(2x+z) + {1\over{4}}d^2z^2 + m_{\pi}^4)} 
\eqno(10) $$      

\noindent               
The remaining integral over the other variables can be calculated by combining 
analytical and numerical techniques, for both the ND and D limits.   

\section{Results}

We estimated first the total neutrino emissivity for the processes $nn\nu{\bar
{\nu}}$ and $pp\nu{\bar{\nu}}$ since they involve the same contribution from the
nuclear potential in both the ND and D cases.  
For comparison we normalize to the results of Thompson, Burrows and 
Horvath (2000) \cite{[TBH20]} valid 
for a constant matrix element and arbitrary degeneracies. 
Denoting their 
result for the $nn$ (or $pp$) neutrino emissivity with $Q_{b}^{nn}$ we find an
{\it under}evaluation of their calculation of a factor of about two as 
compared with our result. More precisely we found

$$ \frac{Q_{\nu}^{nn} - Q_b^{nn}}{Q_b^{nn}} = 1.065  \eqno(11)$$   
     
in the ND limit. Furthermore, this difference is {\it independent} 
of $y$ as long as the ND limit applies 
and $T \, <  \, m_{\pi}^{2}/m$ . Care should be taken in this comparison since instead of sticking 
strictly to the high-momentum limit (a procedure that would 
have required the presence of the "3" 
prefactor), Thompson, Burrows and Horvath (2000) have introduced a 
fudge factor $\xi$ to embrace all 
those effects. In terms of that quantity we obtain $\xi \, \sim \, 2$; or in other words, that the "true" 
bremsstrahlung emissivity is about $2/3$ of its high-momentum limit. 

In the D limit, and adopting the same conditions of Flowers, Sutherland and Bond (1975) we obtain 

$$ \frac{Q_{\nu}^{nn} - Q_b^{nn}}{Q_b^{nn}} = 1.0578  \eqno(12)$$

The neutrino emissivity for the case of the 
$np\nu{\bar{\nu}}$ process can be calculated analogously using an analogous procedure 
to obtain $M^{corr}_{np}$, 
but given that in general 
the chemical potentials of the neutrons and protons 
are different from each other this contribution is not so easily evaluated. 
We shall address the emissivity of 
this process and the general case 
for arbitrary degeneracy of all processes in a forthcoming paper.

\section{Conclusions}

In this work we developed a general method for calculating 
the total emissivity of the neutrino pair production from NN bremsstrahlung, 
taken into account the full contribution of an OPE nuclear potential. 
We have showed that for particular physical conditions, characterized 
by $T  \, < \, 20 \, MeV$, 
the multiple integral appearing in the emissivity formulae can be performed. 
Both in the ND and D regimes we found that the inclusion of the nuclear potential 
contribution produces neutrino emissivities which are $\sim \, 2/3$ of their respective 
high-momentum limits. The method for including the full contribution of a nuclear 
OPE potential in the calculation  
of the neutrino emissivites from NN bresstrahlung is rather general
and allows the computation of all $nn, pp$ and $np$ processes in both ND and D 
limits and also it could be  used to treat other proceeses of astrophysical interest, 
in which NN interaction is important.     

Some key issues still remain to be clarified for a full evaluation of the bremsstrahlung 
neutrino emissivity, the most important perhaps is the interplay between the correlations 
in the dense medium (\cite{[HR95]}, \cite{[BS98]}) and the momentum dependence of 
the nuclear potential. Other refinements (like the inclusion of multipion exchange) are not 
expected to be crucial, at least in the ND or mildly degenerate conditions corresponding 
to the semitransparent post-shock conditions in a SNII, although they may be relevant 
in other situations. Another issue is the corresponence with the recently claimed reduction 
extracted in a model-independent analysis by Hanhart, Phillips and Reddy \cite{[HPR]}. 
We have found that the inclusion of the full OPE potential indeed reduces the emission 
rates in both extreme limits, although a detailed comparison with that work 
has not been attempted here. 

\section{Acknowledgement}

S.Stoica wishes to acknowledge Funda\c c\~ao de Amparo \`a Pesquisa do Estado de 
S\~ao Paulo for financial support to visit the IAG/USP . J.E. Horvath wishes to acknowledge 
the CNPq Agency (Brazil) for partial financial support.


\begin{thebibliography}{99}
\bibitem{[TS00]} See S. Tsuruta, Phys. Rep. {\bf 292}, 1 (1998) for a review. 
\bibitem{[LP76]} D. Lamb and C. Pethick, Astrophys. J., {\bf 209}, L77 (1976).
\bibitem{[TBH20]} T. A. Thompson, A. Burrows and J.E. Horvath, Phys. Rev. C {\bf 62}, 
035802 (2000).
\bibitem{[BS98]} A. Burrows and R. Sawyer, Phys. Rev. C {\bf 58}, 554 (1998).
\bibitem{[HR95]} S. Hannestad and G. Raffelt, Phys. Rev. D {\bf 51}, 6635 
(1995).
\bibitem{[HR98]} S. Hannestad and G. Raffelt, Astrophys. J. {\bf 507}, 339
(1998).
\bibitem{[IWA84]} N. Iwamoto, Phys. Rev. Lett. {\bf 53}, 1198 (1984). 
\bibitem{[BT88]} R. P. Brinkmann and M. S. Turner, Phys. Rev. D {\bf 38}, 2338
(1988).
\bibitem{[SIG97]} G. Sigl, Phys. Rev. D {\bf 56}, 3179 (1997).  
\bibitem{[BUR20]} A. Burrows, T. Young, P. Pinto, R. Eastman and T.A.Thompson,\\ 
Astrophys. J.{\bf 539}, 865 (2000).
\bibitem{[RS98]} G. Raffelt and D. Seckel, Phys. Rev. Lett. {\bf 67}, 2605 
(1998).
\bibitem{[RS95]} G. Raffelt and D. Seckel, Phys. Rev. D {\bf 52}, 1780 (1995).
\bibitem{[RSS96]} G. Raffelt, D. Seckel and G. Sigl, Phys. Rev. D {\bf 54}, 
2784 (1996).
\bibitem{[FM79]} B. L. Friman and O. V. Maxwell, Astrophys. J., {\bf 232}, 541
(1979).
\bibitem{[FSB75]} E. G. Flowers, P. G. Sutherland and J. R. Bond, Phys. Rev. D,
{\bf 12}, 2 (1975).
\bibitem{[BW65]} J. N. Bahcall and R. A. Wolf, Phys. Rev. B {\bf 140}, 1452
(1965).
\bibitem{[BM82]} A. Burrows and T. J. Mazurek, Astrophys. J. {\bf 259}, 330
(1982).   
\bibitem{[BEN74]} A. Benvenuti et al., Phys. Rev. Lett. {\bf 32}, 800 (1974).
\bibitem{[HAS73]} F. J. Hasert et al., Phys. Lett. B {\bf 46}, 138 (1973).
\bibitem{[BA73]} S.J. Barish et al., Phys. Rev. Lett. {\bf 33}, 448 (1973). 
\bibitem{[HPR]} C. Hanhart, D.R. Phillips and S. Reddy, Phys. Lett. B {\bf 499}, 9 
(2001); see also A.E.L. Dieperink, E.N.E. van Dalen, A. Korchin and R. Timmermans, nucl-th/0012073 (2000).
\end{thebibliography}
\end{document}